\documentclass[11pt]{amsart}

\newtheorem{Theorem}{Theorem}
\newtheorem{Corollary}{Corollary}
\newtheorem{Definition}{Definition}

\newtheorem{Lemma}{Lemma}

\newcommand{\N}{\ensuremath{{\textup{I\!N}}}}
\newcommand{\Z}{\ensuremath{{\mathsf{Z\!\!Z}}}}
\newcommand{\R}{\ensuremath{{\textup{I\!R}}}}
\newcommand{\D}{\ensuremath{{\textup{I\!D}}}}
\newcommand{\C}{\ensuremath{{\mathrm{C\hspace{-1.7mm}\rule{0.3mm}{2.6mm}\;\,}}}}
\newcommand{\tr}{\mathrm{trace}\,}

\begin{document}

\title[Bernoulli Discrete Dirac Operator]{Spectral and Localization Properties for the
One-Dimensional Bernoulli Discrete Dirac Operator}

\author{C\'{e}sar R. de Oliveira}
\thanks{CRdeO was partially supported by CNPq (Brazil).}
\address{Departamento de Matem\'{a}tica -- UFSCar, S\~{a}o Carlos, SP,
13560-970 Brazil} \email{oliveira@dm.ufscar.br}
\author{Roberto A. Prado}
\thanks{RAP was supported by CAPES (Brazil).}
\address{Departamento de Matem\'{a}tica -- UFSCar, S\~{a}o Carlos, SP,
13560-970 Brazil\\} \email{rap@dm.ufscar.br}


\begin{abstract} A {\sc 1D} Dirac tight-binding model is considered
and it is shown that its nonrelativistic limit is the {\sc 1D}
discrete Schr\"odinger model. For random Bernoulli potentials
taking two values (without correlations), for typical realizations
and for all values of the mass, it is shown that its spectrum is
pure point, whereas the zero mass case presents dynamical
delocalization for specific values of the energy. The massive case
presents dynamical localization (excluding some particular values
of the energy). Finally, for general potentials the dynamical moments for distinct masses are
compared, especially the massless and massive Bernoulli cases.
\end{abstract}

\maketitle

\

\section{Introduction} Besides the huge amount of mathematical works on spectral problems related
to the one-dimensional Dirac model~\cite{BjD,T}, in physics it has also been used in comparative
studies of relativistic and nonrelativistic electron-lo\-cal\-iz\-ation phenomena~\cite{Ba}, in
relativistic investigations of electrical conduction in disordered systems~\cite{RBa}, in the
construction of supertransparent models with supersymmetric structures~\cite{Sta} and in
relativistic tunnelling problems~\cite{R}.

In this paper a discrete version of the {\sc 1D} Dirac model is discussed, which can be
interpreted as a relativistic version of the well-known tight-binding Schr\"odinger
Hamiltonian (with
$\hbar=1$)
\begin{equation}
\label{Soperator}  (H\psi)_n=-\frac{1}{2m}(\Delta\psi)_n +V_n\psi_n = \frac{1}{2m}(-\psi_{n+1}
-\psi_{n-1} + 2\psi_n) + V_n\psi_n.
\end{equation} The model was first reported in~\cite{deOPr} and this work is its very expanded and
mathematical detailed version. Consider a
particle of mass~$m\ge0$ in the one-dimensional lattice~$\Z$ under the real site potential
$\tilde V=(V_n)$. The proposed {\sc 1D} Dirac tight-binding operator is
\begin{equation}
\label{Doperator}  \D(m,c)=\D_0(m,c)+\tilde V\,Id_2= c\mathcal{B} + mc^2\sigma_3 +\tilde V\,Id_2 ,
\end{equation}  with $c>0$ representing the speed of light,
$$\mathcal{B} =\left(\begin{array}{cc} 0 & d^* \\ d & 0 \\
\end{array}\right),$$
$\sigma_3$ the usual Pauli matrix, $Id_2$ the $2\times2$ identity matrix and ${d}$ the finite
difference operator (a discrete counterpart of the first derivative) defined by \[({d}\psi)_n =
\psi_{n+1} - \psi_{n}.\]  $(d^*\psi)_n =
\psi_{n-1}-\psi_n$ is the adjoint of~$d$ so that $\D_0(m,c)=c\mathcal{B} + mc^2\sigma_3$ is a
bounded self-adjoint operator acting on
$\ell^2(\Z;\C^2)$  and its square is
$${\D_0(m,c)}^2=\left(\begin{array}{cc} c^2dd^*+m^2c^4 & 0 \\ 0 & c^2dd^*+m^2c^4 \\
\end{array}\right).$$ This equality is reminiscent of the relation between momentum
$\vec{p}$ and energy
$E$ in relativistic quantum mechanics~\cite{BjD}, given by
$E^2=c^2\vec{p}^{\hspace{0.05cm} 2}+m^2c^4.$ Denoting by
$\sigma(A)$ the spectrum of a self-adjoint operator $A$, it is well known
that~$\sigma(-\Delta)=[0,4]$, and since $d^*d=dd^*=~-\Delta$,
\[
\sigma(\D_0(m,c))=\left[-c\sqrt{4+m^2c^2},-mc^2\right] \cup
\left[mc^2,c\sqrt{4+m^2c^2}\right].
\] In case the potential~$\tilde V$ is a bounded sequence,
$\D(m,c)$ is also a bounded self-adjoint operator acting on
$\ell^2(\Z;\C^2)$.

It will be shown that the nonrelativistic limit of  the resolvent of the discrete
Dirac operator~(\ref{Doperator}) is the resolvent of discrete Schr\"{o}dinger
operator~(\ref{Soperator}) (when projected on a proper subspace; see
Section~\ref{LimSection}). This is an important support for such Dirac model.

The study of quantum transport depends, of course, on the admitted definitions. In the physics
literature terms like ``extended states'' and ``zero Lyapunov exponents'' have been used to
crudely designate quantum transport. For instance, in~\cite{TC} it was claimed that ``extended
states'' were found in one-dimensional Schr\"odinger systems with off-diagonal randomness, but
in~\cite{SE} it was argued that although the localization length diverges the ``transmission
coefficient'' vanishes as the system size goes to infinity.  Up to recently, in the
mathematical literature pure point spectrum (sometimes with exponentially decaying
eigenfunctions) was considered synonymous of absence of transport. Currently the
transport has been probed via the time behavior of the moments of the position operator, and
in this work this idea will be followed. See ahead for precise definitions and related
comments.

One of the goals of this paper is to study the phenomenon called dynamical localization (in
the sense of time-boundedness of all moments of the position operator) for the Bernoulli-Dirac
model, that is, the model~(\ref{Doperator}) with the site potentials $V_n,\: n\in\Z ,$ being
independent identically distributed Bernoulli random variables taking the values $\pm V,\
V>0.$ In this case it will be shown that almost surely the spectrum of $\D(m,c)$ is pure point
for all values of the mass, the massive case has dynamical localization (excluding some
particular values of the energy for which a more careful analysis is needed) and the zero mass
case presents dynamical delocalization (that is, absence of localization) for specific values
of the energy.

The problem of dynamical localization has been intensively studied during last years,
especially in the case of random discrete and continuous Schr\"odinger
operators (in particular for the Bernoulli-Anderson model, that is, the Schr\"odinger model with
Bernoulli potentials); see~\cite{DeG,FrMSS,GD} and references there in. What one
usually proves is the so-called exponential localization~\cite{An,CKM,DrK}, i.e., pure point
spectrum and exponentially decaying eigenfunctions. On the other hand, it is also known that
exponential localization does not imply dynamical localization~\cite{DelJLS}; it is usually
needed a precise control of the decay of the eigenfunctions, called SULE~\cite{DelJLS,GD}, that
can be obtained through the method of multiscale analysis, a technique set out by
Fr\"ohlich and Spencer~\cite{FrS,FrMSS}.

One motivation for studying dynamical localization for the Bernoulli-Dirac operator comes
from the random dimer model~\cite{DuWP,DeG}, i.e., the Bernoulli-Anderson model with the site
energies $V_n$ assigned for pairs of lattices: $V_{2n}=V_{2n+1}=\pm V$ for all~$n$. This
model almost surely presents pure point spectrum for all values of $V\ne0$~\cite{DeG}.
It was also numerically found in~\cite{DuWP} and rigorously shown
in~\cite{DeG,JSS} the existence of {\sl critical energies} (in the sense of~\cite{JSS}; see
ahead) at which the Lyapunov exponent vanishes; dynamical localization was obtained in~\cite{DeG}
only after projecting onto closed energy intervals not containing such critical energies. Despite
the similarity between the transfer matrices of the two models, it is not immediate the
adaptation of the localization (delocalization) results to the Bernoulli-Dirac model and each
step needs to be verified; here, many points will not be detailed when they follow exactly the
same lines of their Schr\"odinger counterpart.

With respect to nontrivial quantum transport, probed via dynamical
delocalization (unbounded moments of the position operator), it was found in random
polymer models~\cite{JSS} and in random palindrome models~\cite{CaO} (both including the
important random dimer model), due to existence of critical energies~\cite{JSS}. Recently, for 
{\sc 1D} discrete Schr\"odinger operators, Damanik, S\"ut\H o and Tcheremchantsev~\cite{DST} have
developed a general method which allows one to derive quantum dynamical lower bounds from upper
bounds on the growth of norms of transfer matrices, and they applied this method to
some substitution, Sturmian and prime models, among others.  Damanik, Lenz and
Stolz~\cite{DLS} presented an extension of this method to {\sc 1D} continuous Schr\"odinger
operators, with application to the continuous Bernoulli-Anderson model. Another method to obtain
quantum transport from upper bounds on transfer matrix  was lately developed by Germinet,
Kiselev and Tcheremchantsev~\cite{GKT}, with application to  Schr\"odinger operators with random
decaying potentials, providing  new examples of Schr\"odinger operator with point spectrum and
nontrivial quantum transport. 

In the zero mass case, the one-dimensional Bernoulli-Dirac model presented here has pure point
spectrum and nontrivial quantum transport for potentials with no
correlations nor decaying properties (see Section~\ref{LocSection}). This phenomenon does not
take place in the corresponding Schr\"odinger tight-binding model~\cite{GD}, and this also
motivates the interest to better understanding the Dirac case. Presumably, this tight-binding
model is the simplest one presenting such phenomenon.

Now the localization results for the Bernoulli-Dirac model will be briefly summarized. By using
as the main tool a particular form of Furstenberg Theorem~(Lemma~\ref{FurstenbergTeor} ahead),
it is shown (see Theorems \ref{DinamicTeor1}, \ref{DinamicTeor2} and
\ref{CriticalTeor}) that the Lyapunov exponent~$\gamma_m(E)$ is strictly positive for the
energies $E\in\sigma(\D(m,c)),$ except for: $E=\pm V,\ V\in(0,c], \ V\neq c/\sqrt{2},$ in the
case $m=0$; and  if~$m\geq 0$ for  $\big(E_V=0,\ V=c\sqrt{2+m^2c^2}\big)$ and the four
energy-potential pairs
$\big(E_V=\pm c\sqrt{2+m^2c^2}\pm c/\sqrt2,\ V=c/\sqrt 2\big)$.

For all energies~$E$ for which $\gamma_m(E)>0$,  a initial estimate for localization
(Lemma~\ref{I.E.L.}) and the Wegner's estimate (Lemma~\ref{W}) will be checked; by adapting the
method multiscale analysis~\cite{DrK,FrMSS,GD} to this model, it will be shown (see Theorems
\ref{DinamicTeor1} and
\ref{DinamicTeor2}) that for typical realizations the spectrum of $\D(m,c)$ is pure point and
the corresponding eigenfunctions are semi-uniformly exponentially localized
(SULE)~\cite{DelJLS,GD}. This and the results of~\cite{GD} (properly adapted to $\D(m,c)$)
imply dynamical localization.

In the massless $(m=0)$ case, the values $E=\pm V$ with $V\in(0,c],\ V\neq c/\sqrt{2},$ are
critical energies for the operator $\D(0,c)$ and this implies (almost surely) upper boundedness
for the transfer matrices in the vicinity of these energies (see Lemmas
\ref{Lim.Unif.Matrices} and \ref{Lim.Matrices}). By adapting the ideas of \cite{JSS}
(see also \cite{DST})
to $\D(0,c)$ it will follow (see Theorem~\ref{DelocTeor}) that for an initial spinor
$\Psi$ well localized in space, there is $0<C_q<\infty$ such that
\[\int_0^{\infty}\frac{1}{T} \ {e^{-2t/T}} M_{\Psi}^{(q)}(0,t) dt
\geq C_q T^{q-1/2}, \] for almost all realization of the potential (or exponent $q-1$ instead
of $q-1/2$ for every realization), where ($X$ is the usual position operator)
\[
M_{\Psi}^{(q)}(m,t):=\langle e^{-i\D(m,c)t}\Psi,|X|^q
e^{-i\D(m,c)t}\Psi \rangle,
\]  i.e., there is nontrivial
quantum transport despite the absence of a continuous component in the spectrum of $\D(0,c)$.

In the case of the set of pairs  $\big(E_V=\pm c\sqrt{2+m^2c^2}\pm c/\sqrt2,\
V=c/\sqrt 2\big)$ and $\big(E_V=0,\ V=c\sqrt{2+m^2c^2}\big)$ it is shown (see Theorem
\ref{CriticalTeor}) that the Lyapunov exponent $\gamma_m$ vanishes, but it was not
possible to give an answer about dynamical localization for them. Nevertheless, for these cases
there is a general dynamical upper bound (in fact valid for all potentials $\tilde V$)
established in Theorem~\ref{upperNovas}.

For distinct masses $m, m'\geq 0$, but $m$ close to~$m'$, it is expected that the moments
$M_{\Psi}^{(q)}(m,t)$ follow closely the moments $M_{\Psi}^{(q)}(m',t)$ (both with the same
potential), at least for a small period of time. The final result to be reported is an
inequality confirming such expectative; by making using of Duhamel's formula, it will be shown
(see Theorem~\ref{CompTeor}) that, for the initial state~$\Psi$ with only one nonzero
component, there exists $K_q> 0$ so that, for all $t>0$,
\[ \left|M^{(q)}_{\Psi}(m,t)-M^{(q)}_{\Psi}(m',t)\right|
\leq K_q |m-m'|c^2 \: t^{q+2}.\] In particular, for the Bernoulli-Dirac model this relation with
$m'=0$ gives quantitatively an estimate of how, for small times, the dynamics of the localized
regime follows the delocalized one (see also Corollary~\ref{corolAs} in
Section~\ref{CompDinamicSection}).

This paper is organized as follows: In Section~\ref{LimSection} the nonrelativistic limit for
the discrete Dirac model~(\ref{Doperator}) is discussed. In Section~\ref{LocSection} the
results about spectral properties of such model, dynamical localization (delocalization) and a
dynamical upper bound for moments are presented, whose proofs appear in
Section~\ref{ProofsSection}. In Section~\ref{ToolsSection} some
tools used in those proofs are collected. Finally, in the Section~\ref{CompDinamicSection} the
dynamical moments with different masses are compared; in particular the dynamics of the massless
and massive Dirac-Bernoulli cases.

\

\section{Nonrelativistic Limit}
\label{LimSection} In this section consider $\D(m,c)$ with~$m>0$
fixed and~$c$ as a parameter. For simplicity, $\D(c)$
will denote~$\D(m,c)$, which is supposed to be self-adjoint with (real) potential ${\tilde V}$.

The nonrelativistic limit means~$c$ going to infinity, and since the
rest energy $mc^2$ is a purely relativistic quantity, (as usual) it must be subtracted
before taking this limit. The norm
convergence of the resolvent operators
$\left(\D(c)-mc^2-z\right)^{-1}$, for $z\in \C \backslash\R$ will be considered.  $\Lambda$
below is the projector onto the subspace of ``positive energies,'' and so $\Lambda H_\infty$ 
corresponds to the Schr\"odinger operator~(\ref{Soperator}). It is interesting to compare the
approach presented here with the one in~\cite{deOPr}.

\

\begin{Theorem}\label{LimCor} If $z\in \C \backslash\R$,
then
$$\lim_{c \to \infty}\left(\D(c)-mc^2-z \right)^{-1}=
\Lambda\left(H_{\infty}-z\right)^{-1},$$  where $\Lambda=\displaystyle\frac{1}{2}\left(Id_2+\sigma_3\right)$ \ and
\ $H_{\infty}=\displaystyle\frac{\mathcal{B} ^2}{2m}+\tilde V\Lambda,$  and the limit is in
the norm of bounded operators.
\end{Theorem}

\

\begin{Lemma}\label{Res.Teor}
If $z\in \C \backslash\R$, then
\begin{equation}\label{ResEq1}
\left(\D(c)-mc^2-z \right)^{-1}= \left(\Lambda+\frac{c\mathcal{B}
+z}{2mc^2}\right)S(c) \left(Id+\tilde V\frac{c\mathcal{B}
+z}{2mc^2}S(c)\right)^{-1} ,
\end{equation} where $Id$ is the identity operator and
\begin{equation}\label{ResEq2}
S(c)=\left(H_{\infty}-z-\frac{z^2}{2mc^2}\right)^{-1}=
\left(Id-\frac{z^2}{2mc^2}\left(H_{\infty}-z\right)^{-1}
\right)^{-1}\left(H_{\infty}-z\right)^{-1}.
\end{equation}
\end{Lemma}

\begin{proof} Note that
$$\left(\D_0(c)+mc^2+z\right)\left(\D_0(c)-mc^2-z\right)=
c^2\mathcal{B} ^2-2mc^2z-z^2.$$ Hence
\begin{eqnarray}\label{ResEq3}
\left(\D_0(c)-mc^2-z\right)^{-1}& = &\frac{\D_0(c)+mc^2+z}{2mc^2}
\left(\frac{\mathcal{B} ^2}{2m}-z-\frac{z^2}{2mc^2}\right)^{-1} \\
& = &\left(\Lambda+\frac{c\mathcal{B} +z}{2mc^2}\right)S_0
\nonumber
\end{eqnarray} with
$S_0=\left(\displaystyle\frac{\mathcal{B}
^2}{2m}-z-\frac{z^2}{2mc^2}\right)^{-1}$ \hspace{-0.3cm}. \ On the
other hand, by using the operator relation
$$(A+B)^{-1}=(Id-A^{-1}B)^{-1}A^{-1}$$  with \
$A=\displaystyle\frac{\mathcal{B}
^2}{2m}-z-\displaystyle\frac{z^2}{2m}$ \ and \ $B=\tilde V\Lambda$, one
obtains 
\begin{equation}\label{ResEq4} S(c)=S_0\left(Id+\tilde V\Lambda
S_0\right)^{-1}.
\end{equation} Therefore, by (\ref{ResEq3}) and (\ref{ResEq4})
it is found that
\begin{eqnarray*}
\left(\D(c)-mc^2-z\right)^{-1}& = &
\left(\D_0(c)-mc^2-z\right)^{-1}
\left(Id+\tilde V\left(\D_0(c)-mc^2-z\right)^{-1}\right)^{-1} \\ & = &
\left(\Lambda+\frac{c\mathcal{B} +z}{2mc^2}\right)S_0
\left(Id+\tilde V\Lambda S_0+\tilde V\frac{c\mathcal{B}
+z}{2mc^2}S_0\right)^{-1} \\ & = &
\left(\Lambda+\frac{c\mathcal{B} +z}{2mc^2}\right)S(c)
\left(Id+\tilde V\frac{c\mathcal{B} +z}{2mc^2}S(c)\right)^{-1}.
\end{eqnarray*}
\end{proof}

\

\begin{proof}{\bf (Theorem~\ref{LimCor})} Since $\left(H_{\infty}-z\right)^{-1}$ is
bounded for
$z\in \C \backslash\R$ \ and
$$\left\|\frac{z^2}{2mc^2}\left(H_{\infty}-z\right)^{-1}\right\|<1$$
for~$c$ sufficiently large, one can expand
\begin{equation}\label{Expan1}
S(c)=\sum_{n=0}^{\infty}\left(\frac{z^2}{2mc^2}
\left(H_{\infty}-z\right)^{-1}\right)^{n}\left(H_{\infty}-z\right)^{-1},
\end{equation}
where the sum is convergent in the operator norm.

For any fixed $z\in\C \backslash\R$ \ and $c$ sufficiently large,
$$\left\|T(c):= {\tilde V} \frac{c\mathcal{B} +z}{2mc^2}S(c)\right\|<1$$
and so
\begin{equation}\label{Expan2}
\left(Id+T(c)\right)^{-1}=\sum_{n=0}^{\infty}\left(-T(c)\right)^n
\ .
\end{equation}
Replacing (\ref{Expan1}) and (\ref{Expan2}) into (\ref{ResEq1})
one obtains the expansion
\[
\left(\D(c)-mc^2-z\right)^{-1}= \sum_{n=0}^{\infty}
\frac{R_n(z)}{c^n}
\] with 
\[
R_0(z)=\Lambda\left(H_{\infty}-z\right)^{-1}  , 
\]
\[R_1(z)=\Lambda\left(H_{\infty}-z\right)^{-1}\frac{\mathcal{B}
}{2m}+ \frac{\mathcal{B}
}{2m}\left(H_{\infty}-z\right)^{-1}\Lambda   , 
\] and so on, and the sum is convergent in the operator norm. The result
then follows.
\end{proof}

\

\section{Localization Results}
\label{LocSection} Consider the family of Dirac operators
\begin{equation} \label{DirOper} \D_{\omega}(m,c)=
\left(\begin{array}{cc} mc^2 & cd^* \\ cd & -mc^2 \\ \end{array}
\right)+ V_{\omega}Id_2, \quad \omega\in\Omega=\{-V,V\}^{\Z},
\end{equation} on $\ell^2(\Z;\C^2)$, where $V_{\omega}(n),\ n\in\Z,$ are
i.i.d.\ Bernoulli random variables taking the values $\pm V,\
V>0,$ with common (nontrivial) probability measure $\mu$ and
product measure
$\textbf{P}=\prod_{n\in\Z}\mu\left(V_{\omega}(n)\right).$  Let
$P^\omega_{I,m}$ be the spectral projector of $\D_{\omega}(m,c)$
onto the interval~$I\subset\R$.

Denote by $\delta_n^{\pm}$ the elements of the canonical  position
basis of $\ell^2(\Z;\C^2)$, for which all entries are
$\left(\begin{array}{c} 0 \\ 0 \\ \end{array}\right)$ except at
the $n$th entry, which is given by $\left(\begin{array}{c} 1 \\ 0
\\ \end{array}\right)$ and $\left(\begin{array}{c} 0 \\ 1 \\
\end{array}\right)$
   for the superscript  indices~$+$ and~$-$, respectively. If
$\Psi=\left(\begin{array}{c} \psi^+ \\ \psi^- \end{array}\right)$
is a solution of the eigenvalue equation
\[ (\D_{\omega}(m,c)-E)\Psi=0,
\] then it is  simple to check that
\[ \left( \begin{array}{c}\psi^+(n+1) \\ \psi^-(n) \\
\end{array}\right) = T^{V_{\omega}(n)}_m(E) \left(
\begin{array}{c}\psi^+(n) \\ \psi^-(n-1) \\
\end{array}
   \right), \] with
\[ T^V_{m}(E)=\left(
\begin{array}{cc} 1+\displaystyle\frac{m^2c^4-(E-V)^2}{c^2} &
\displaystyle\frac{mc^2 +E-V}{c}\\ \\
\displaystyle\frac{mc^2-E+V}{c} & 1 \\
\end{array} \right).\]
The transfer matrix from site $k$ to site $n$ is introduced by
\[ T^{\omega}_{m}(E;n,k)=T^{V_{\omega}(n-1)}_m(E)T^{V_{\omega}(n-2)}_m(E)
\cdots T^{V_{\omega}(k)}_m(E) \ , \  \ n>k \] and
$T^{\omega}_m(E;n,n)=Id_2.$ \ For $q>0$, let $|X|^q$ be the moment
of order~$q$ of the position operator on $\ell^2(\Z;\C^2)$, i.e.,
$$|X|^q \Psi=\sum_n |n|^q \left(\langle
\delta_n^+,\Psi \rangle \delta_n^+ +\langle \delta_n^-,\Psi
\rangle \delta_n^-  \right).$$

\

\begin{Definition}\label{dinamicdef} The operator $\D_{\omega}(m,c)$ is
dynamically localized on a spectral interval~$I$ if for all $q>0$
and for all exponentially decaying initial state $\Psi\in
\ell^2(\Z;\C^2)$
$$\sup_t M_{\Psi,I,\omega}^{(q)}(m,t):=\sup_t \ \langle
P^\omega_{I,m}e^{-i\D_{\omega}(m,c)t}\Psi,|X|^q
P^\omega_{I,m}e^{-i\D_{\omega}(m,c)t}\Psi \rangle < \infty $$
$\textbf{P}$ almost surely ($\textbf{P}$-a.s.). Otherwise
$\D_{\omega}(m,c)$ is dynamically delocalized on~$I$. If
$I=\sigma(\D_{\omega}(m,c))$, then $M_{\Psi,I,\omega}^{(q)}(m,t)$
will be denoted by $M_{\Psi,\omega}^{(q)}(m,t)$.
\end{Definition}

\

It is important to notice that although the Dirac operator acts on spinors, its eigenvalue
equation, in the transfer matrix form, looks exactly like the equation for a one-dimensional
Schr\"odinger operator acting on scalar valued functions, with the transfer matrix being in
$SL(2,\R)$. Hence the methods used in studies of the usual one-dimensional Anderson model, as
Furstenberg's Theorem, can be applied for this Dirac model; see Sections~\ref{ToolsSection}
and~\ref{ProofsSection}. 

The localization  results are gathered in the following set of
theorems.

\

\begin{Theorem}\label{DinamicTeor1} Let
$\left(\D_{\omega}(m,c)\right)_{\omega\in\Omega}$ be as
in~(\ref{DirOper}) and $V \in  (0,c], V\ne c/\sqrt{2}.$ Then,
$\textbf{P}$ almost surely, the Lyapunov exponent
\[\gamma_m(E)=\lim_{n\to\infty} \frac{1}{|n|} \ln
\left \|T^{\omega}_m(E;n,1)\right \| \] exists, is independent of
$\omega$, and
\begin{itemize}
\item[(i) ] \begin{itemize}
\item[(i.1) ]$\gamma_m(E\neq\pm V)>0$ for $m\geq0,$
\item[(i.2) ]$\gamma_m(E=\pm V)>0$ for $m>0,$
\item[(i.3) ]$\gamma_0(E=\pm V)=0.$
           \end{itemize}
\item[ (ii) ] Let $m\ge0$; then $\textbf{P}$-a.s.\
$\sigma(\D_{\omega}(m,c))$ is pure point.
\item[ (iii) ] \begin{itemize}
\item[(iii.1)] Let $m>0$. Then $\textbf{P}$-a.s.\ the operator
$\D_{\omega}(m,c)$ is dynamically localized on its spectrum.
\item[(iii.2)] For any closed interval~$I\subset
\sigma(\D_{\omega}(0,c)),$ with $\pm V\not\in I,$ the operator
$\D_{\omega}(0,c)$ is dynamically localized on~$I$.
               \end{itemize}
\end{itemize}
\end{Theorem}

\

\begin{Theorem}\label{DinamicTeor2} Let
$\left(\D_{\omega}(m,c)\right)_{\omega\in\Omega}$ be as in
(\ref{DirOper}), $m\ge0$ and $V>c,\ V\neq c\sqrt{2+m^2c^2}$. Then,
$\textbf{P}$-a.s.\  the spectrum of \ $\D_{\omega}(m,c)$ is pure
point and this operator is dynamically localized on its spectrum.
\end{Theorem}

\

\begin{Theorem}\label{CriticalTeor} Let
$\left(\D_{\omega}(m,c)\right)_{\omega\in\Omega}$ be as in
(\ref{DirOper}), $m\ge0$ and $V=c/\sqrt{2}$ (respectively
$V=c\sqrt{2+m^2c^2}$). Then the same conclusions of
Theorem~\ref{DinamicTeor1} (resp.\ Theorem~\ref{DinamicTeor2})
hold except at the four possibilities of energies $E_V=\pm
c\sqrt{2+m^2c^2}\pm c/\sqrt{2}$ (resp.\ $E_V=0$). (The point is that $E_V=\pm
c\sqrt{2+m^2c^2}\pm c/\sqrt{2}$ (resp.\ $E_V=0$) are energies so that $\gamma_m(E_V)=0.$)
\end{Theorem}

\

For the next result it is convenient to use the average
dynamical moments
\begin{equation}\label{DefA}
A_{\Psi,\omega}^{(q)}(m,T):= \int_0^{\infty}\frac{1}{T} e^{-2t/T}
M_{\Psi,\omega}^{(q)}(m,t)\ dt,
\end{equation}
defined for $m\ge0$ and $T>0$. The main reason for working with this kind of Laplace transform
average is relation~(\ref{DelocGreen}) ahead.

\

\begin{Theorem} [massless case] \label{DelocTeor} Let
$\left(\D_{\omega}(0,c)\right)_{\omega\in\Omega}$ be as in
(\ref{DirOper}) and $V\in (0,c]$, $V\ne c/\sqrt{2}.$ Then, for $q>0$ and
$\Psi$ with only one nonzero component, there exists
$0<C_q(\omega)<\infty$ such that, for $T>0$,
\begin{itemize}
\item[(i)]$A_{\Psi,\omega}^{(q)}(0,T) \geq  C_q(\omega) T^{q-1/2}
\quad\quad  \textbf{P}-a.s.,$
\item[(ii)]$A_{\Psi,\omega}^{(q)}(0,T) \geq  C_q(\omega) T^{q-1}$
\quad\quad  for every $\omega$,
\end{itemize}
i.e., $\D_{\omega}(0,c)$ is not dynamically localized on its
spectrum.
\end{Theorem}

\

The following theorem establishes  very general upper bounds for the dynamical moments of the
position operator; notice that it holds for any potential sequence~$\tilde V$ and is not
restricted to the Bernoulli case. 

\

\begin{Theorem} \label{upperNovas} Let
$\D(m,c)$ be as in~(\ref{Doperator}), $m\ge0$ and $\Psi$ with only one nonzero
component (so in the domain of $|X|^q$ for all~$q>0$). Then for any $q\in\N$ there exists
$0<K_{q}(\tilde{V},m,c)<\infty$ such that
\[M_{\Psi}^{(q)}(m,t) \leq K_{q}(\tilde{V},m,c) \, \,t^{q},\quad t\ge1.
\]
\end{Theorem}

\

\noindent{\it Remark.} It is possible to adjust the constant $K_q$ so that the above upper bound
holds for $t\ge\varepsilon$ for any given $\varepsilon >0$, instead of just $t\ge1$. Since
$M_{\Psi}^{(q)}(m,t)\ge M_{\Psi}^{(q')}(m,t)$ for $q\ge q'$, it is evident that
$M_{\Psi}^{(q)}(m,t) \le K_{\lceil q\rceil}(\tilde{V},m,c) \, \,t^{\lceil q\rceil}$ for real~$q$.

\

\section{Tools}
\label{ToolsSection} In this section some tools and notations that
will be used in the proofs of the results presented in
Section~\ref{LocSection} are collected. For studying the
positivity of the Lyapunov exponent $\gamma_m ,\ m\geq 0,$ the
following particular form of Furstenberg Theorem~\cite{BoL} will
be used:

\

\begin{Lemma} \label{FurstenbergTeor} Let $\mathcal{G}_m(E)$ be
the smallest closed subgroup of $SL(2,\R)$ generated by the
matrices $T_m^{V}(E)$ and $T_m^{-V}(E).$ Then $\gamma_m(E)>0$ if
\begin{itemize}
\item $\mathcal{G}_m(E)$ is not compact, and
\item there
is no probability measure on $P(\R^2)$ (the set of all the directions
of $\R^2$) that is invariant under the action of
$\mathcal{G}_m(E)$, which is equivalent to the statement: the
orbit\ $\mathcal{G}_m(E)\cdot{\tilde x}:= \{T\cdot {\tilde x},\
T\in \mathcal{G}_m(E)\}$ of each direction ${\tilde x}\in P(\R^2)$
contains at least three elements.
\end{itemize}
\end{Lemma}
\

If $L>0,\ n\in\Z$, consider the finite subset of $\Z$
\[\Lambda_L(n)=\bigg\{k\in\Z : \ |k-n|\leq \frac{L}{2}\bigg\}\] with
boundary
\[\partial\Lambda_L(n)=\{(k,k'):\ k\in\Lambda_L(n),\
k'\not\in\Lambda_L(n),\ |k-k'|=1\}.\] Denote by
$\D_{\omega}^{\Lambda_L(n)}(m,c)$ the operator $\D_{\omega}(m,c)$
restricted to $\ell^2(\Lambda_L(n);\C^2)$ with zero boundary
conditions outside $\Lambda_L(n)$.

The matrix elements of an operator $\mathcal{O}$ on
$\ell^2(\Z;\C^2)$ are given by
$$\mathcal{O}_{nk}=\left(\begin{array}{cc}
\langle\delta_n^+,\mathcal{O}\delta_k^+\rangle &
\langle\delta_n^+,\mathcal{O}\delta_k^-\rangle \\ \\
\langle\delta_n^-,\mathcal{O}\delta_k^+\rangle &
\langle\delta_n^-,\mathcal{O}\delta_k^-\rangle \\
\end{array}\right)$$ with ``norm''
\[\|\mathcal{O}_{nk}\|^2=
|\langle\delta_n^+,\mathcal{O}\delta_k^+\rangle|^2+
|\langle\delta_n^+,\mathcal{O}\delta_k^-\rangle|^2+
|\langle\delta_n^-,\mathcal{O}\delta_k^+\rangle|^2+
|\langle\delta_n^-,\mathcal{O}\delta_k^-\rangle|^2.\]

Now two important results required for the multiscale
analysis are described. The first one is the Wegner's estimate,
adapted from~\cite{CKM} to the discrete Dirac operator (details
will be omitted, since they are long and very similar to the Schr\"odinger case):

\

\begin{Lemma}\label{W} Let $\D_{\omega}(m,c)$ be as in (\ref{DirOper})
and~$I$  a compact energy interval. For any $\theta\in(0,1)$ and
$\tau
>0$ there exist $L_0=L_0(I,\theta,\tau,m)>0$ and
$a=a(I,\theta,\tau,m)>0$ such that
\[\textbf{P}\ \bigg\{\omega:
{\rm
dist}\left(E,\sigma\big(\D_{\omega}^{\Lambda_L(0)}(m,c)\big)\right)
\leq e^{-\tau L^{\theta}}\bigg\} \leq e^{-aL^{\theta}}\] for all
$E\in I $ and $L\geq L_0.$
\end{Lemma}

The second result is the initial estimate for localization,
adapted from~\cite{DrK} (details omitted):

\

\begin{Lemma}\label{I.E.L.} Let $\D_{\omega}(m,c)$ be as in
(\ref{DirOper}), $\epsilon >0$ and $\theta\in (0,1).$ For each
$E_0\in\R,$ $E_0 \not\in
\sigma\big(\D_{\omega}^{\Lambda_L(0)}(m,c)\big)$ with
$\gamma_m(E_0)>\epsilon$, there exist
$L_0=L_0(E_0,\epsilon,\theta,m)>0$ and
$r=r(E_0,\epsilon,\theta,m)>0$ such that
\[\textbf{P}\ \left\{\omega:\
\left\|\left(\D_{\omega}^{\Lambda_L(0)}(m,c)-E_0\right)^{
-1}_{0k}\right\| \leq e^{-(\gamma_m(E_0)-\epsilon)L/2} \  \
\forall \ k\in
\partial\Lambda_L(0) \right\}\geq \ 1-e^{-rL^{\theta}}\] for all $L\geq
L_0.$
\end{Lemma}

\

In order to obtain dynamical localization from the multiscale
analysis, the
following properties of $\D_{\omega}(m,c)$ are useful: \\

\noindent{\bf (P1)} With respect to the spectral measure of
$\D_{\omega}(m,c)$, almost every energy is a generalized
eigenvalue, i.e., with polynomially bounded eigenvector (see~\cite{Be,Si}).\\

\noindent{\bf (P2)} If $E \not\in
\sigma\big(\D_{\omega}^{\Lambda_L(n)}(m,c)\big)$ and $\Psi\in
\ell^2(\Z;\C^2)$ so that $\D_{\omega}(m,c)\Psi=E\Psi,$ then
\begin{eqnarray*}
\Psi(n) &=& -\left(\D_{\omega}^{\Lambda_L(n)}(m,c)-E
\right)^{-1}_{nl_1} \left(\begin{array}{cc} 0 & c \\ 0 & 0 \\
\end{array}\right) \Psi(l_1 -1) \\ & &
-\left(\D_{\omega}^{\Lambda_L(n)}(m,c)-E
\right)^{-1}_{nl_2} \left(\begin{array}{cc} 0 & 0 \\ c & 0 \\
\end{array}\right) \Psi(l_2 +1),
\end{eqnarray*} with $\{(l_1,l_1-1), (l_2,l_2+1)\}
= \partial\Lambda_L(n).$ \\

\

Property {\bf (P2)} follows after defining the {\sl boundary
operator} $\mathcal F_{\Lambda_L(n)}$ by its matrix elements
\[
\left(\mathcal F_{\Lambda_L(n)} \right)_{jk}= \left\{
\begin{array}{rrr} -\left(\begin{array}{cc} 0 & c \\ 0 & 0 \\
\end{array}\right) \quad\;\;  \mathrm{if}\
j-1=k,\  j\in \Lambda_L(n),\  k\notin \Lambda_L(n); \\
-\left(\begin{array}{cc} 0 & 0 \\ c & 0 \\ \end{array}\right) \quad\;\;
\mathrm{if}\  j+1=k,\  j\in
\Lambda_L(n),\  k\notin \Lambda_L(n); \\
\quad\left(\begin{array}{cc} 0 & 0 \\ 0 & 0 \\ \end{array}\right)
\hfill  \mathrm{ otherwise,}\hfill
\end{array}
\right.
\]
noting that $l^2(\Z;\C^2)=l^2(\Lambda_L(n);\C^2)\oplus
l^2(\Z\setminus \Lambda_L(n);\C^2)$ and
\[
\D_{\omega}(m,c)=\D_{\omega}^{\Lambda_L(n)}(m,c)+\D_{\omega}^{\Z\setminus
\Lambda _L(n)}(m,c)-{\mathcal F}_{\Lambda_L(n)}.
\]

\

In the zero mass case ($m=0$) the operators $\D_{\omega}(0,c),\
\omega\in\Omega,$ presents critical energies $E_V=\pm V$ for
$V\in(0,c]\ , V\neq c/\sqrt{2}$, as defined in~\cite{JSS}, since
either $T_0^{V}(V)=Id_2$ and $T_0^{-V}(V)$ is elliptic (that is, $|\tr
\ T_0^{-V}(V)|<2$) or $T_0^{-V}(-V)=Id_2$ and $T_0^{V}(-V)$ is
elliptic. Thus there exists a real invertible matrix $Q$ such that
$$Q\ T_0^{\pm V}(E_V)\ Q^{-1}=\left(\begin{array}{cc}
\cos(\eta_{\pm}) & -\sin(\eta_{\pm}) \\
\sin(\eta_{\pm}) & \cos(\eta_{\pm}) \\
\end{array}\right).$$
Since the eigenvalues of this matrix are $e^{i\eta_{\pm}}$ and
$e^{-i\eta_{\pm}}$, for both of the above cases one has $\eta_+
-\eta_- \neq k\pi,\ k\in\Z$ (a condition required in~\cite{JSS}).
By using the modified Pr\"{u}fer variables, phase shifts, oscillatory
sums, large deviation estimates as in~\cite{JSS}, one obtains the
following result

\

\begin{Lemma}[massless case] \label{Lim.Unif.Matrices}
Let $\lambda>0$ be arbitrary. Then there are $b>0$ and $C<\infty$
such that for every $N\in\N,$ there exists a set
$\Omega_N(\lambda)\subset\Omega$ with
$\textbf{P}\left(\Omega_N(\lambda)\right)\leq Ce^{-bN^{\lambda}}$
and
\[\big\|T_0^{\omega}(E;n,k)\big\| \leq C \]
for all $\omega\in\Omega\backslash\Omega_N(\lambda)$, \ $0\leq
k\leq n\leq N$ and $E\in
[E_V-N^{-\lambda-1/2},E_V+N^{-\lambda-1/2}] .$
\end{Lemma}

\

On the other hand, since $\|Q\ T_0^{\pm V}(E_V)\ Q^{-1}\|=1$,
expanding $T_0^{\pm V}(E_V+\epsilon)$ into powers of $\epsilon$
one obtains \[\|Q\ T_0^{\pm V}(E_V+\epsilon)\ Q^{-1}\|\leq
1+a|\epsilon|\] for $|\epsilon|\leq\delta$,\ $0<a<\infty$, and one
deduces the following

\

\begin{Lemma}[massless case] \label{Lim.Matrices}
For $\delta >0$ there exists  $C<\infty$ such that for all \
$n,k\in\Z$ and $E\in [E_V-\delta,E_V+\delta],$
\[\big\|T_0^{\omega}(E;n,k)\big\| \leq Ce^{C\delta|n-k|}. \]
\end{Lemma}

\

An inductive argument shows that, for $\zeta\in\C$ and $m\geq 0$,
\begin{equation}\label{PER}
T_m^{\omega}(E+\zeta;n,k)=T_m^{\omega}(E;n,k)-\zeta\sum_{l=k}^{n-1}
T_m^{\omega}(E+\zeta;n,l+1)\ S_{\zeta}^{\omega}(E;l)\
T_m^{\omega}(E;l,k) ,\end{equation} where
$$S_{\zeta}^{\omega}(E;l)=\displaystyle\frac{\zeta}{c^2}
\left(\begin{array}{cc} 1 & 0 \\ 0 & 0 \\
\end{array}\right)+\displaystyle\frac{1}{c}
\left(\begin{array}{cc} \displaystyle\frac{2}{c} (E-V_\omega(l)) &
-1 \\ 1 & 0 \\
\end{array}\right).$$ 

\

 Now, for $z\in \C\backslash\R$ and $m\geq 0$,
introduce the two-components Green's function 
\[\left(\begin{array}{c} G_{m,\omega}^+(z;n) \\  G_{m,\omega}^-(z;n)
\end{array}\right)= \left(\begin{array}{c}
\left\langle \delta_n^+,
\left(\D_{\omega}(m,c)-z\right)^{-1}\delta_0^+ \right\rangle
\\  \left\langle \delta_n^-,
\left(\D_{\omega}(m,c)-z\right)^{-1}\delta_0^+ \right\rangle
\end{array}\right), \] so that
\[(\D_{\omega}(m,c)-z)
\left(\begin{array}{c} G_{m,\omega}^+(z;n) \\  G_{m,\omega}^-(z;n)
\end{array}\right)= \delta_0^+(n) \ . \]
By using transfer matrices, one obtains for $n\leq 0$,
\begin{equation} \label{MatrGreen1}
\left(\begin{array}{c} G_{m,\omega}^+(z;n) \\
G_{m,\omega}^-(z;n-1)
\end{array}\right)= T_m^{\omega}(z;n,0)
\left(\begin{array}{c} G_{m,\omega}^+(z;0) \\
G_{m,\omega}^-(z;-1)
\end{array}\right)
\end{equation} and for $n\geq 1$,
\begin{equation} \label{MatrGreen2}
\left(\begin{array}{c} G_{m,\omega}^+(z;n) \\
G_{m,\omega}^-(z;n-1)
\end{array}\right)= T_m^{\omega}(z;n,1)
\left(\begin{array}{c} G_{m,\omega}^+(z;1) \\  G_{m,\omega}^-(z;0)
\end{array}\right). \end{equation}

For $z=E+i/T \ (T>0)$ and $m\geq 0$, it is also valid the
following identity (adapted from Lemma~3.2 in~\cite{KKL}):
\begin{equation} \label{DelocGreen}
A_{\delta_0^+,\omega}^{(q)}(m,T)=\frac{1}{2\pi
T}\sum_{n\in\Z}|n|^q
\int_{\R}(|G_{m,\omega}^+(z;n)|^2+|G_{m,\omega}^-(z;n)|^2)\ dE.
\end{equation}

\

\section{Localization Proofs}
\label{ProofsSection} In this section the proofs of
Theorems~\ref{DinamicTeor1}~-~\ref{upperNovas} are presented.

\

\begin{proof} {\bf (Theorems \ref{DinamicTeor1} and~\ref{DinamicTeor2})}

The strategy of the proof is based on reference~\cite{DeG}, where the random dimer Schr\"odinger
operator was studied. Since for the discrete Dirac operator there are the particular role played
by the mass and some different possibilities for the transfer matrices, a rather detailed proof
will be presented. The idea is to show that given
$\epsilon>0$, $I\subset\sigma\left(\D_{\omega}(m,c)\right)$ a
compact energy interval  not containing the excluded $V$ values,
then for each $0<\gamma<\gamma_m(I):=\inf \{\gamma_m(E): E\in I\}$
there exist a constant $C(\omega,\epsilon,\gamma)$ and, for each
eigenfunction
${\varphi}_{j,\omega}=\left(\begin{array}{c} {\varphi}_{j,\omega}^+ \\
{\varphi}_{j,\omega}^- \\
\end{array}\right)$ with energy $E_{j,\omega}\in I,$ a ``center''
$z_{j,\omega}\in\Z,$ such that
\begin{equation}\label{EqSULE}
\|{\varphi}_{j,\omega}(n)\| \leq
C(\omega,\epsilon,\gamma)e^{\gamma
|z_{j,\omega}|^{\epsilon}}e^{-\gamma |n-z_{j,\omega}|},\quad
\forall n\in\Z.
\end{equation} If $\Psi$ decays exponentially with rate
$\theta_0 >0$ and if $q>0$, it is known that~(\ref{EqSULE})
(that is, SULE condition) implies the  existence of a constant
$C_{\Psi}(m,I,\omega)$ so that
\[
\sup_t M_{\Psi,I,\omega}^{(q)}(m,t) \leq C_{\Psi}(m,I,\omega),
\quad \textbf{P}-a.s.,
\] i.e., $\D_{\omega}(m,c)$ is dynamically localized
on~$I$ (see Section~2 in~\cite{GD}).

To prove (ii) and (\ref{EqSULE}), it is sufficient to show strict
positivity of the Lyapunov exponent, because in this case
Lemmas~\ref{W} and~\ref{I.E.L.} hold. By using the multiscale
analysis \cite{DrK} together with (P1) and (P2), one can then follow
the proof of Theorem~3.1 in~\cite{GD} (properly adapted to
$\D_{\omega}(m,c)$) to obtain (ii) and (\ref{EqSULE}) (details will be omitted).

Now the proof of~(i). It follows from Furstenberg and Kesten
Theorem~\cite{BoL} that, $\textbf{P}$-a.s.\ the Lyapunov exponent
$\gamma_m$ exists and is independent of $\omega$.

Consider first the energies $E\neq \pm V$ and it will be proven that
$\gamma_m(E\neq \pm V)>0$ for all $m\geq 0$ and for all
$E\in\sigma(\D_{\omega}(m,c))$. Let $\mathcal{G}_m(E)$ be as in
the Lemma~\ref{FurstenbergTeor}. Put $\alpha=E-V$, $\beta=E+V$
and rename $T_m^{V}(E)=T_m^{(\alpha)}$,
$T_m^{-V}(E)=T_m^{(\beta)}$. In the present case $\alpha\neq 0$
and $\beta\neq 0$.

Since the problem is symmetric in $\alpha$ and $\beta$, the proof
is reduced to the study of three cases:
\begin{itemize}
\item[\bf a)] $T_m^{(\alpha)}$ and $T_m^{(\beta)}$ are elliptic $(|{\tr}
T_m^{(\alpha)}|<2,\ |{\tr} T_m^{(\beta)}|<2)$;

\item[\bf b)] $T_m^{(\alpha)}$ is parabolic $(|{\tr} T_m^{(\alpha)}|=2)$;

\item[\bf c)] $T_m^{(\alpha)}$ is hyperbolic $(|{\tr} T_m^{(\alpha)}|>2)$.
\end{itemize}
Note that in cases {\bf b)} and {\bf c)} the group
$\mathcal{G}_m(E)$ is not compact. \\

\noindent {\bf Case a)} Since $T_m^{(\alpha)}$ and
$T_m^{(\beta)}$ are both elliptic, then $|\alpha|, |\beta| \in
(mc^2,c\sqrt{4+m^2c^2})$. In this case such matrices do not commute. Since
the operator
\[ T_m^{(\alpha)}T_m^{(\beta)}\
(T_m^{(\alpha)})^{-1}(T_m^{(\beta)})^{-1} \] built from two noncommuting
elliptic elements is hyperbolic, it follows that
$\mathcal{G}_m(E)$ is not compact. Moreover, note that
\[{\tr}\big(T_m^{(\alpha)}\big)^2= \frac{{\alpha}^4}{c^4}-\left
(2m^2+\frac{4}{c^2}\right) {\alpha}^2 + m^2c^2(4+m^2c^2)+2 \]
\big(analogous for $T_m^{(\beta)}$\big). Hence, if ${\alpha}^2\neq
2c^2+m^2c^4$ or ${\beta}^2\neq 2c^2+m^2c^4$, then $T_m^{(\alpha)}$
and $\big(T_m^{(\alpha)}\big)^2$ or $T_m^{(\beta)}$ and
$\big(T_m^{(\beta)}\big)^2$ are elliptic. Since elliptic elements
have no fixed points in $P(\R^2)$, it follows that for any $\tilde
x \in P(\R^2), \ \mathcal{G}_m(E)\cdot{\tilde x}$ contains at
least the three elements $\tilde x, \ T_m^{(\alpha)}\cdot\tilde x, \
\big(T_m^{(\alpha)}\big)^2\cdot\tilde x$ or $\tilde x, \
T_m^{(\beta)}\cdot\tilde x, \ \big(T_m^{(\beta)}\big)^2\cdot\tilde
x$. Therefore, by Lemma~\ref{FurstenbergTeor},  $\gamma_m(E)>0$.
If, on the other hand, ${\alpha}^2= 2c^2+m^2c^4$ and ${\beta}^2=
2c^2+m^2c^4$, then $E=0$ and $V=c\sqrt{2+m^2c^2}$, which is one of
the excluded pairs described
in Theorem~\ref{CriticalTeor}. 

\

\

\noindent {\bf Case b)} Suppose $T_m^{(\alpha)}$ is parabolic,
that is, $|\alpha|=mc^2$ or $|\alpha|=c\sqrt{4+m^2c^2}$. First the possibility $\alpha= mc^2$ will
be discussed (the case
$\alpha= -mc^2$ is similar). In this case
\[T_m^{(\alpha)}=
\left(\begin{array}{cc} 1 & 2mc \\ 0 & 1 \\
\end{array}\right), \quad \mbox{and so} \quad
\big(T_m^{(\alpha)}\big)^n=
\left(\begin{array}{cc}
1 & 2nmc \\ 0 & 1 \\
\end{array}\right) .\] Denote by $\{e_1,e_2\}$ the canonical basis of $\R^2$. By taking a vector
$x=x_1e_1+x_2e_2$, and setting $\tilde x$
for its direction, one concludes that $\lim_{n\to\infty}
\big(T_m^{(\alpha)}\big)^n\cdot\tilde x = \tilde e_1$. If $\nu$ is
a probability measure that is invariant under the action of
$\mathcal{G}_m(E)$, and if $f\in C_0^\infty (P(\R^2))$, by
Lebesgue's dominated convergence Theorem one has
\[f(\tilde e_1)=\lim_{n\to\infty}\int
f\left(\big(T_m^{(\alpha)}\big)^n\cdot\tilde x\right)\
d\nu(\tilde x)\ .\] This means that $\nu=\delta_{\tilde e_1}$. But
the matrix $T_m^{(\beta)}$ does not leave invariant the direction
$\tilde e_1$ since
\[T_m^{(\beta)} \ e_1= \bigg(1+\frac{m^2c^4-{\beta}^2}{c^2}\bigg)e_1+
\frac{-\beta+mc^2}{c}\ e_2 \  \  \  \mbox{and} \ \ \beta\neq mc^2
\ .\] Thus it is proven that there is no invariant measure under the
action of $\mathcal{G}_m(E)$. Therefore, by
Lemma~\ref{FurstenbergTeor} one gets $\gamma_m(E)>0$.

Consider now the possibility $\alpha=c\sqrt{4+m^2c^2}$ (the case
$\alpha=-c\sqrt{4+m^2c^2}$ is similar). In this case an eigenvector
of
$$T_m^{(\alpha)}= \left(\begin{array}{cc} -3 & mc+\sqrt{4+m^2c^2} \\ \\
mc-\sqrt{4+m^2c^2} & 1 \\ \end{array}\right) $$ is given by
$v_1=\bigg(\displaystyle\frac{mc+\sqrt{4+m^2c^2}}{2},1\bigg)$.
Picking $v_2=\bigg(\displaystyle\frac{-mc+\sqrt{4+m^2c^2}}{2},-1\bigg)$
a vector orthogonal to $v_1$, the matrix $T_m^{(\alpha)}$ in the basis
$\{v_1,v_2\}$ is $$\left(\begin{array}{cc} -1 &
-4-m^2c^2+mc\sqrt{4+m^2c^2} \\
\\ 0 & -1 \\ \end{array}\right).$$ Repeating the previous calculation
for this case, one obtains $\nu=\delta_{\tilde v_1}$. But
$T_m^{(\beta)}$ does not leave invariant the direction $\tilde v_1$
except for $\beta= 0$ or $\beta=c\sqrt{4+m^2c^2}=\alpha$, which are
excluded since the first condition yields $E=-V$ and the second
one $V=0$. Thus it is proven that there is no invariant measure and, by
Lemma~\ref{FurstenbergTeor},
\ $\gamma_m(E)>0$. 

\

\

\noindent {\bf Case c)} Suppose now that $T_m^{(\alpha)}$ is
hyperbolic (so $|\alpha|<mc^2$ or $|\alpha|>c\sqrt{4+m^2c^2}$). It is
sufficient to study the orbit of the eigendirections of
$T_m^{(\alpha)}$, namely
$$e_m^{\epsilon}=\left(\begin{array}{c}
{\alpha}^2-m^2c^4+\epsilon\sqrt{({\alpha}^2-m^2c^4)({\alpha}^2-m^2c^4
-4c^2)}
\\ \\ 2c(\alpha -mc^2) \\ \end{array}\right), \quad \epsilon=\pm
1.$$ If $T_m^{(\beta)}$ is hyperbolic then the orbit of
$e_m^{\epsilon}$ is infinite. Hence $\gamma_m(E)>0$ by
Lemma~\ref{FurstenbergTeor}. If $T_m^{(\beta)}$ is parabolic, it is
 again  case~b). Finally, suppose that $T_m^{(\beta)}$ is
elliptic. If $T_m^{(\beta)} \tilde e_m^{\epsilon} \neq \tilde
e_m^{-\epsilon}$, then $T_m^{(\beta)} \tilde e_m^{\epsilon}$ cannot
belong to the eigendirections of $T_m^{(\alpha)}$ and its orbit is
infinite. Hence $\gamma_m(E)>0$ by Lemma~\ref{FurstenbergTeor}. If
$T_m^{(\beta)} \tilde e_m^{\epsilon}= \tilde e_m^{-\epsilon}$, then
simple calculations lead to the equations
\[\bigg(1+m^2c^2-\frac{{\beta}^2}{c^2}\bigg)
({\alpha}^2-m^2c^4+\epsilon u)= 4(m^2c^4-\beta\alpha)+
({\alpha}^{2}-m^2c^4-\epsilon u)\] with $\epsilon=\pm 1$ and
$u=\sqrt{({\alpha}^2-m^2c^4)({\alpha}^2-m^2c^4-4c^2)}\neq 0$. It
implies ${\beta}^{2}=2c^2+m^2c^4$ and $\alpha=\beta \pm
c\sqrt{2}$, which means $V=c/\sqrt{2}$ and $E=\pm
c\sqrt{2+m^2c^2}-c/\sqrt{2}$. The symmetric case where one assumes
that $T_m^{(\beta)}$ is hyperbolic leads naturally to
${\alpha}^{2}=2c^2+m^2c^4$ and $\beta=\alpha \pm c\sqrt{2}$, which
means  $V=c/\sqrt{2}$ and $E=\pm
c\sqrt{2+m^2c^2}+c/\sqrt{2}$. Those are excluded pairs that will
be discussed in the proof of
Theorem~\ref{CriticalTeor}. \\

Consider now the energy $E=V$ (the case $E=-V$ is analogous). Note
that $\alpha= 0$ and $\beta= 2V$. First the case $m>0$ will be
discussed. The two possible transfer matrices are
$$T_m^{(\alpha)}=\left(\begin{array}{cc} 1+m^2c^2 & mc \\ mc & 1 \\
\end{array}\right) \ \ \ \mbox{and} \ \ \
T_m^{(\beta)}=\left(\begin{array}{cc}
1+\displaystyle\frac{m^2c^4-4V^2}{c^2} &
\displaystyle\frac{mc^2+2V}{c} \\ \\
\displaystyle\frac{mc^2-2V}{c} & 1 \\
\end{array}\right).$$ Observe that $T_m^{(\alpha)}$ and $T_m^{(\beta)}$ do not
commute, and that $T_m^{(\alpha)}$ is hyperbolic. It is enough to study
this case for $\beta=c\sqrt{4+m^2c^2}$ \big($T_m^{(\beta)}$ is
parabolic\big). The eigendirections of $T_m^{(\alpha)}$ are
$$e_m^\delta=\left(\begin{array}{c}
\displaystyle\frac{mc+\delta\sqrt{4+m^2c^2}}{2} \\ \\ 1 \\
\end{array}\right), \quad \delta=\pm 1. $$ The matrices $T_m^{(\alpha)}$ and
$T_m^{(\beta)}$ in the basis $\{e_m^1,e_m^{-1}\}$ are given,
respectively, by
$$\left(\begin{array}{cc} \lambda_1 & 0 \\ 0 & \lambda_{-1} \\
\end{array}\right) \quad \mbox{and} \quad
\left(\begin{array}{cc} -1 & 4+m^2c^2-mc\sqrt{4+m^2c^2} \\ \\ 0 & -1 \\
\end{array}\right) ,$$ with
$$\lambda_1 \lambda_{-1}=\bigg(1+\frac{m^2c^2}{2}+
\frac{mc\sqrt{4+m^2c^2}}{2}\bigg)\bigg(1+\frac{m^2c^2}{2}
-\frac{mc\sqrt{4+m^2c^2}}{2}\bigg)=1.$$

Suppose that $T_m^{(\alpha)}$ occurs with probability $0<p<1$ and
$T_m^{(\beta)}$ occurs with probability $1-p$. Denote by
$n_\alpha$ (resp.\ $n_\beta$) the number of times that
$T_m^{(\alpha)}$ (resp.\ $T_m^{(\beta)}$) occurs in the product
$T_m^{\omega}(E;n,1)$. Supposing, without loss of generality, that
$T_m^{V_{\omega}(1)}(E)=T_m^{(\alpha)}$, one has
\[T_m^{\omega}(E;n,1)=\left(\begin{array}{cc} \lambda_1^{n_\alpha}
& C_n
P(\lambda_1,\lambda_{-1}) \\  \\ 0 & \lambda_{-1}^{n_\alpha} \\
\end{array}\right)\]
$\textbf{P}$-a.s., where $C_n$ is a constant and
$P(\lambda_1,\lambda_{-1})$ is a polynomial in $\lambda_1$ and
$\lambda_{-1}$. Thus,
\[ \Big\|T_m^{\omega}(E;n,1)\Big\|\geq \left\|\left(\begin{array}{c}
\lambda_1^{n_\alpha} \\ \\ 0  \\ \end{array}\right)\right\|=
\lambda_1^{n_\alpha} , \ \lambda_1>1, \] and therefore
\textbf{P}-\mbox{a.s.}
\[
\gamma_m(E=V)=\lim_{n\to\infty}
\frac{1}{|n|}\ln\|T^{\omega}_m(E;n,1)\| \geq (\ln
\lambda_1)\lim_{n\to\infty}\frac{n_\alpha}{|n|}=(\ln \lambda_1)p
> 0.
\]

Now the case $m=0$ will be treated. In this case
\[T_0^{(\alpha)}= Id_2 \quad \mbox{and} \quad
T_0^{(\beta)}=\left(\begin{array}{cc}
1-4V^2/c^2 & 2V/c \\ \\ -2V/c & 1 \\ \end{array}\right).\] One then finds
$$\lim_{n\to\infty}\big\|\big(T_0^{(\beta)}\big)^n\big\|^{1/n}=1$$
if $V\in (0,c]$ and
 $$ \lim_{n\to\infty}\big\|\big(T_0^{(\beta)}\big)^n\big\|^{1/n}>1$$
if $V>c.$ Hence, if  $V\in (0,c]$, $V\ne c/\sqrt 2$,
$$\gamma_0(E=V)=\lim_{n\to\infty} \frac{n_\beta}{|n|} \ln
\big\|\big(T_0^{(\beta)}\big)^{n_\beta}\big\|^{1/n_\beta}=(1-p)\ln
1= 0,$$ and $\gamma_0(E=V)>0$ \ if \ $V>c$.
\end{proof}

\

\begin{proof}{\bf (Theorem~\ref{CriticalTeor})}

By analyzing the proof of Theorems~\ref{DinamicTeor1}
and~\ref{DinamicTeor2} observe that for $V=c/\sqrt{2}$ (resp.\
$V=c\sqrt{2+m^2c^2}$) one has $\gamma_m(E_V\neq\pm
c\sqrt{2+m^2c^2}\pm c/\sqrt{2})>0$ (resp.\
$\gamma_m(E_V\neq~0)>0$) and then the same conclusions of
Theorem~\ref{DinamicTeor1} (resp.\ Theorem~\ref{DinamicTeor2})
hold. It remains to show that $\gamma_m$ vanishes at the 
pairs $\big(V=c\sqrt{2+m^2c^2},\ E_V=0\big)$ and
$\big(V=c/\sqrt{2},\ E_V=\pm c\sqrt{2+m^2c^2}\pm c/\sqrt{2}\
\big)$. Note that in all cases $E_V\in\sigma(\D_{\omega}(m,c))$ \
$\textbf{P}$-a.s.\ .

First it will be treated the case $\big(V=c/\sqrt{2},\
E_V=-c\sqrt{2+m^2c^2} -c/\sqrt{2}\ \big)$ (the others three
excluded cases with $V=c/\sqrt{2}$ are similar). In this case one has  $\beta=-c\sqrt{2+m^2c^2}$
and $\alpha=\beta -c\sqrt{2}$. The eigenvectors of $T_m^{(\alpha)}$ are given by
$$\left(\begin{array}{c} \displaystyle\frac{2c-\sqrt{2}\beta+\epsilon
\sqrt{4c^2+2m^2c^4-2\sqrt{2}c\beta}}{\beta-c\sqrt{2}-mc^2} \\ \\ 1
\\ \end{array}\right), \quad \epsilon=\pm 1,$$ and by looking at the
matrices in the basis given by these two vectors, the study is reduced to
products of matrices of the following two types:
$$\left(\begin{array}{cc}
\lambda_+ & 0 \\0& \lambda_-
\\ \end{array}\right) \ \ \mbox{and} \ \
\left(\begin{array}{cc} 0 & \kappa_{-} \\
\kappa_{+} & 0 \\ \end{array}\right)$$ with $\lambda_+\lambda_-=1$
\ and \ $\kappa_{+} \kappa_{-}=-1$, where
$$\lambda_{\pm}=-1+\frac{\sqrt{2}\beta}{c}\pm
\frac{\sqrt{4c^2+2m^2c^4-2\sqrt{2}c\beta}}{c}$$ and
$$\kappa_{\pm}=\Big(-\frac{\beta}{c}+mc\Big)
\Big(\frac{2c-\sqrt{2}\beta\pm
\sqrt{4c^2+2m^2c^4-2\sqrt{2}c\beta}}{\beta-c\sqrt{2}-mc^2}\Big)+1.$$
Moreover,
$$\big(T_m^{(\beta)}\big)^2=\left(\begin{array}{cc} -1 & mc+\beta /c
\\ \\ mc-\beta /c & 1
\\ \end{array}\right)^2 = -Id_2 \ .$$ Therefore the proof that
$\gamma_m\big(E_V=-c\sqrt{2+m^2c^2} -c/\sqrt{2},\ V=c/\sqrt{2}\
\big)=0$ is analogous to the Schr\"{o}dinger case (see the proof of
Theorem~2.4 in~\cite{DeG}).

Now consider the excluded case $\big(V=c\sqrt{2+m^2c^2},\
E_V=0\big)$. In this case ${\alpha}^2={\beta}^2=2c^2+m^2c^4$.
Since $\alpha\neq\beta$ (otherwise $V=0$), then $\alpha=-\beta=\pm
c\sqrt{2+m^2c^2}$. Noting that
$\big(T_m^{(\alpha)}\big)^2=\big(T_m^{(\beta)}\big)^2=-Id_2$ and
$T_m^{(\alpha)} T_m^{(\beta)}$ is hyperbolic, the proof that
$\gamma_m\big(E_V=~0,\ V=c\sqrt{2+m^2c^2}\ \big)=0$ is again similar to the corresponding
Schr\"{o}dinger case (see the proof of Theorem~2.4 in~\cite{DeG}).
\end{proof}

\

\begin{proof}{\bf (Theorem~\ref{DelocTeor})}

(i) It is sufficient to prove the Theorem for $\Psi=\delta_0^+$.
Given $\lambda >0$, there exists $b'>0$ and by
Lemma~\ref{Lim.Unif.Matrices} there are $b>0$ and $C<\infty$ such
that, by applying Lemma~\ref{Lim.Unif.Matrices} for $N=[b'T]$,
together with the relation~(\ref{PER}) for $m=0$ and $\zeta=i/T$,
one concludes that there exists a set \
$\Omega_N(\lambda)\subset\Omega$ with
$\textbf{P}(\Omega_N(\lambda))\leq Ce^{-bN^{\lambda}}$ and
\begin{equation}\label{Lim.MT}
\big\|T_0^{\omega}\big(E+i/T;n,1\big)\big\| \leq C
\end{equation}
for all $\omega\in\Omega\backslash\Omega_N(\lambda)$, $1\leq n\leq
N$ and $E\in I_{V}=[E_V-N^{-\lambda-1/2},E_V+N^{-\lambda-1/2}]$.

Supposing that
\[|G_{0,\omega}^+(E+i/T;1)|^2+|G_{0,\omega}^-(E+i/T;0)|^2\geq
B_1(\omega)>0, \] it follows from~(\ref{MatrGreen2}) and
\[\|T_0^{\omega}(E+i/T;n,1)^{-1}\|=\|T_0^{\omega}(E+i/T;n,1)\|\]
that
\begin{equation}\label{MatrGreen3}
\max\{|G_{0,\omega}^+(E+i/T;n)|^2,|G_{0,\omega}^-(E+i/T;n-1)|^2
\}\geq
\displaystyle\frac{B_1(\omega)}{2\|T_0^{\omega}(E+i/T;n,1)\|^2} \
.
\end{equation}
Thus, replacing (\ref{Lim.MT}) and (\ref{MatrGreen3}) into
(\ref{DelocGreen}), \textbf{P}-a.s.\ one has
\[A_{\delta_0^+,\omega}^{(q)}(0,T)\geq \frac{1}{2\pi T}\hspace{-0.15cm}
\sum_{0\leq n\leq[b'T]}\hspace{-0.2cm} n^q
\int_{I_{V}}\hspace{-0.15cm}\frac{B_1(\omega)}{2C^2} \ dE \geq
B_q(\omega)T^{q}N^{-\lambda-1/2} \geq C_q(\omega)
T^{q-1/2-\lambda}\] for some constant $C_q(\omega)>0$.

If, on the other hand,
\[|G_{0,\omega}^+(E+i/T;0)|^2+|G_{0,\omega}^-(E+i/T;-1)|^2\geq
B_2(\omega)>0,\] then one gets this estimate in the same way, but
based on~(\ref{MatrGreen1}) instead of~(\ref{MatrGreen2}). Since
$\lambda >0$ is arbitrary, this finishes the proof.

(ii) It follows from the above arguments by using Lemma~\ref{Lim.Matrices}.
\end{proof}

\

\begin{proof} {\bf (Theorem~\ref{upperNovas})} 

The arguments will be a variation of~\cite{Si2}. Define the operator 
\[
p:= i \left[\D (m,c),X\right]= ci \left(\begin{array}{cc} 0 & -d^*-1 \\ d+1 & 0 \\
\end{array}\right) 
\]
with $[\cdot,\cdot]$ denoting the commutator. Note that $p$ is self-adjoint and bounded. Set
\[
X(t)=e^{i\D(m,c)t}Xe^{-i\D(m,c)t}\quad{\rm and}\quad p(t)=e^{i\D(m,c)t}pe^{-i\D(m,c)t},
\]
so that 
\[
\frac{d}{dt}X(t)=i\left[\D (m,c),X(t)\right]=e^{i\D(m,c)t} i\left[\D
(m,c),X\right]e^{-i\D(m,c)t}=p(t).
\]
Hence
\[
X(t)=X+\int_0^t p(s)\, ds.
\]
Using this relation, the boundedness $\|p(t)\|=\|p\|<\infty$ for all~$t$,  Cauchy-Schwarz
inequality, and keeping only the dominant terms for large~$t$, it follows that for $t\ge1$ and
$q\in\N$, there exists
$C_q(\tilde V,m,c)>0$ so that
\begin{eqnarray*}
M_{\Psi}^{(q)}(m,t) &=& \langle\Psi,|X(t)|^q\Psi\rangle \\ &\le& C_q(\tilde
V,m,c)\int_0^t\cdots\int_0^t
\langle\Psi,p(s_1)\cdots p(s_q)\Psi\rangle\, ds_1\cdots ds_q \\ &\le & C_q(\tilde
V,m,c) \|p\|^q\,t^q = K_q(\tilde V,m,c)\, t^q.
\end{eqnarray*}
\end{proof}

\section{Dynamical Comparison}
\label{CompDinamicSection}

The aim of this section is to compare the dynamical moments
$M_{\Psi}^{(q)}(m,t)$, as in Definition~\ref{dinamicdef}, for
different masses and general potentials, in particular for the
massless and massive Bernoulli cases. 

\

\begin{Theorem}\label{CompTeor} Let $\D(m,c)$ and $\D(m',c)$ be
Dirac operators on $l^2(\Z,\C^2)$ defined as in~(\ref{Doperator})
with the same potential~$\tilde V$, and let $\Psi$ be with only one nonzero
component. Given $T>0$, there exists a constant $B_q>0$ so that
\begin{equation} \label{CompDinamic}
\sup_{0\leq t\leq
T}\Big|M_{\Psi}^{(q)}(m,t)-M_{\Psi}^{(q)}(m',t)\Big| \leq B_q\,
|m-m'|c^2\, T^{q+2}.
\end{equation}
\end{Theorem}

\begin{proof}\ Observe that for $m=m'$ the result is immediate.
Suppose $m\neq m'$. For the proof it will be assumed that $m>0,\ m'=0$ and $\Psi=\delta_0^+$ (the
case  $m>0,\ m'>0$ and $\Psi$ as in the hypotheses is similar).

For fixed $\alpha >0$ consider the Banach space
\[B_{\alpha}:=\left\{\Phi\in l^2(\Z,\C^2):\ \|\Phi\|_{\alpha}=
\sup_{k\in\Z} e^{\alpha |k|}\Big(|\langle \delta_k^+,\Phi \rangle
|+ |\langle \delta_k^-,\Phi \rangle |\Big)<\infty \right\}.\]
Since $\D(m,c)$ is a bounded operator on $B_{\alpha}$, it follows
that
\begin{eqnarray}\label{EqEvolucao} |\langle
\delta_n^+,e^{-i\D(m,c)t}\delta_0^+ \rangle |+ |\langle
\delta_n^-,e^{-i\D(m,c)t}\delta_0^+ \rangle | &\leq&
\|e^{-i\D(m,c)t}\delta_0^+\|_{\alpha} \ e^{-n\alpha} \\ &\leq&
e^{-n\alpha+t \|\D(m,c)\|_{\alpha}}. \nonumber
\end{eqnarray}

For $k\in\N$ denote by $X^k$ the restriction of the position
operator~$X$ to the set $\{n\in\Z:\ |n|\leq k\}$ and by
$M_{\delta_0^+}^{(q),k}(m,t)$ the corresponding dynamical moments.
Then, for all times $t\leq \displaystyle\frac{\alpha \
k}{2\|\D(m,c)\|_{\alpha}}$, using (\ref{EqEvolucao}) one has
$$\left|M_{\delta_0^+}^{(q)}(m,t)-M_{\delta_0^+}^{(q),k}(m,t)
\right|=
$$
\begin{eqnarray}\label{CompDinamic1}
=\sum_{|n|>k}|n|^q \left(\left|\langle \delta_n^+,e^{-i\D(m,c)t}\delta_0^+
\rangle \right|^2+ \left|\langle
\delta_n^-,e^{-i\D(m,c)t}\delta_0^+ \rangle \right|^2\right)\nonumber \\
\leq\hspace{-0.1cm} C_1(q) k^q
e^{-k\alpha+2t\|\D(m,c)\|_{\alpha}}\leq \ C_1(q) k^q .
\end{eqnarray} Furthermore, it follows by DuHamel's formula that
\[M_{\delta_0^+}^{(q),k}(m,t)-M_{\delta_0^+}^{(q),k}(0,t)=\]\[
-i\int_0^t \left\langle \delta_0^+,e^{i\D(m,c)t}|X^k|^q
e^{-i\D(m,c)(t-s)}\left(\D(m,c)-\D(0,c)\right)e^{-
i\D(0,c)s}\delta_0^+ \right\rangle \: ds\]\[ +i\int_0^t \left\langle
\delta_0^+,e^{i\D(m,c)(t-s)}(\D(m,c)-\D(0,c)) e^{i\D(0,c)s}|X^k|^q
e^{-i\D(0,c)t}\delta_0^+\right\rangle \: ds.
\] Hence, for $t\leq \displaystyle\frac{\alpha \
k}{2\|\D(m,c)\|_{\alpha}}$, using~(\ref{EqEvolucao}), the fact of the operator
 $e^{i\D(m,c)t}$ on~$\ell^2(\Z;\C^2)$ be unitary and Cauchy-Schwarz, it
is found that
\begin{eqnarray}\label{CompDinamic2}
\Big|M_{\delta_0^+}^{(q),k}(m,t)-M_{\delta_0^+}^{(q),k}(0,t)\Big|&\leq&
C_2(q) mc^2 k^{q+1}t\ e^{-k\alpha+t\|\D(m,c)\|_{\alpha}} \\ &\leq&
C_2(q) mc^2 \frac{\alpha}{2\|\D(m,c)\|_{\alpha}}\
k^{q+2}.\nonumber
\end{eqnarray} Thus, by (\ref{CompDinamic1}) and
(\ref{CompDinamic2}),
\[\left|M_{\delta_0^+}^{(q)}(m,t)-M_{\delta_0^+}^{(q)}(0,t)\right|\leq B_q
mc^2 \frac{\alpha}{2\|\D(m,c)\|_{\alpha}}\ k^{q+2} , \] for all
times $t\leq \displaystyle\frac{\alpha \
k}{2\|\D(m,c)\|_{\alpha}}$. \\

Now, for each $T>0$ choose $k$ to be the smallest integer such
that
$$k\geq \frac{2\|\D(m,c)\|_{\alpha}}{\alpha}\ T.$$ Therefore, for all
$t\leq T$,
\[\Big|M_{\delta_0^+}^{(q)}(m,t)-M_{\delta_0^+}^{(q)}(0,t)\Big| \leq B_q mc^2
T^{q+2}.\]
\end{proof}

With respect to the Bernoulli-Dirac model~(\ref{DirOper}), the
relation (\ref{CompDinamic}) with $m>0$ and $m'=0$ gives an
estimate of how, for small times and/or sufficiently small mass,
the dynamics of the localized regime follows the delocalized one.
In terms of the average dynamical moments
$A_{\Psi,\omega}^{(q)}(m,T)$ defined in~(\ref{DefA}) one has

\

\begin{Corollary}\label{corolAs}  Let
$\left(\D_{\omega}(m,c)\right)_{\omega\in\Omega}$ be as in
(\ref{DirOper}), $m\geq 0$ and $V\in (0,c],V\ne c/\sqrt{2},$ and
let $\Psi$ be with only one nonzero component. Then, for each
$q>0$, $\textbf{P}$-a.s.\ there is $\tilde C_{q,\omega}>0$ so that
\[
\left|1-
\frac{A_{\Psi,\omega}^{(q)}(m,t)}{A_{\Psi,\omega}^{(q)}(0,t)}\right|
\leq \tilde C_{q,\omega}\, mc^2 t^{5/2},\quad t>0.
\]
Notice that the power exponent on the r.h.s.\ of this expression does not
depend on~$q$.
\end{Corollary}
\begin{proof}
By Theorem~\ref{CompTeor} with $m>0$ and $m'=0$, it follows that
for all~$t>0$
\[
\left|A_{\Psi,\omega}^{(q)}(m,t)-A_{\Psi,\omega}^{(q)}(0,t)\right|
\leq \Gamma(q+3) C_{q,\omega}\, mc^2 t^{q+2},
\]
with~$\Gamma$ the usual gamma function. By
Theorem~\ref{DelocTeor}(i)\ there is $0<B_q(\omega)<\infty$ such
that
\[A_{\Psi,\omega}^{(q)}(0,t) \geq B_q(\omega) t^{q-1/2}
\quad\quad \textbf{P}-a.s.,\] and the result follows with
$\tilde C_{q,\omega}=\Gamma(q+3) C_{q,\omega}/B_q(\omega).$
\end{proof}

\

\noindent{\it Remark.} By using Theorem~\ref{upperNovas} one gets (for $q\in\N$)
\[
\left| M_{\Psi}^{(q)}(m,t) - M_{\Psi}^{(q)}(m',t)\right| \leq \tilde{K}_{q}(\omega,m,m',c) \,
\,t^{q},
\] but with no expression for the constant $\tilde{K}_{q}(\omega,m,m',c)$. The price paid for the
explicit dependence on the masses and light speed~$c$ in Theorem~\ref{CompTeor} is the larger
exponent $q+2$ instead of just~$q$. In the same way, the exponent 5/2 in Corollary~\ref{corolAs}
could be replaced by 3/2, but with no precise dependence of the resulting multiplicative constant
on $m$ and~$c$.


\clearpage



\begin{thebibliography}{99}

\bibitem{An} P. W. Anderson, Absence of Diffusion in Certain Random
Lattices, Phys. Rev. {\bf 109} (1958) 1492--1505.

\bibitem{Ba} C. Basu, C. L. Roy, E. Maci\'a, F. Dom\'{\i}nguez-Adame
and A. S\'anchez, Localization of Relativistic Electrons in a
One-Dimensional Disordered System, J. Phys. A {\bf 27} (1994)
3285--3291.

\bibitem{Be} J. M. Berezanskii, Expansions in Eigenfunctions of
Self-Adjoint Operators, Providence: Amer. Mat. Soc., 1968.

\bibitem{BjD} S. D. Bjorken and J. D. Drell, Relativistic Quantum
Mechanics, McGraw-Hill, New York, 1965.

\bibitem{BoL} P. Bougerol and J. Lacroix, Products of Random Matrices
with Applications to  Schr\"odinger Operators, Birkh\"auser,
Boston, 1985.

\bibitem{CKM} R. Carmona, A. Klein and F. Martinelli, Anderson
Localization for Bernoulli and other Singular Potentials, Commun.
Math. Phys. {\bf 108} (1987) 41--66.

\bibitem{CaO} T. O. Carvalho and C. R. de Oliveira, Critical Energies in
Random Palindrome Models, J. Math. Phys. {\bf 44} (2003) 945--961.

\bibitem{DLS} D. Damanik, D. Lenz and G. Stolz, Lower Transport
Bounds for One-Dimensional Continuum Schr\"odinger Operators,
Preprint (2004).

\bibitem{DST} D. Damanik, A. S\"{u}t\H o and S. Tcheremchantsev,
Power-Law Bounds on Transfer Matrices and Quantum Dynamics in One
Dimension II, J. Funct. Anal. {\bf 216} (2004) 362--387.

\bibitem{DeG} S. De Bi\`evre and F. Germinet, Dynamical Localization for
Random Dimer Schr\"odinger Operator, J. Stat. Phys. {\bf 98}
(2000) 1135--1148.

\bibitem{deOPr} C. R. de Oliveira and R. A. Prado, Dynamical Delocalization for the {\sc 1D}
Bernoulli Discrete Dirac Operator, J. Phys. A: Math. Gen. {\bf 38} (2005) L115--L119.

\bibitem{DelJLS} R. del Rio, S. Jitomirskaya, Y. Last and B. Simon,
Operators with Singular Continuous Spectrum IV: Hausdorff
Dimensions, Rank One Perturbations and Localization, J. d'Analyse
Math. {\bf 69} (1996) 153--200.

\bibitem{DuWP} D. H. Dunlap, H.-L. Wu,  and P. W. Phillips, Absence of
Localization in a Random Dimer Model, Phys. Rev. Lett. {\bf 65}
(1990) 88--91.

\bibitem{FrS} J. Fr\"ohlich and T. Spencer, Absence of Diffusion with
Anderson Tight Binding Model for Large Disorder or Low Energy,
Commun. Math. Phys. {\bf 88} (1983) 151--184.

\bibitem{FrMSS} J. Fr\"ohlich, F. Martinelli, E. Scoppola and T.
Spencer, Constructive Proof of Localization in the Anderson Tight
Binding Model, Commun. Math. Phys. {\bf 101} (1985) 21--46.

\bibitem{GD} F. Germinet and S. De Bi\`evre, Dynamical Localization for
Discrete and Continuous Random Schr\"odinger Operators, Commun.
Math. Phys. {\bf 194} (1998) 323--341.

\bibitem{GKT} F. Germinet, A. Kiselev and S. Tcheremchantsev,
Transfer Matrices and Transport for {\sc 1D} Schr\"odinger Operators
with Singular Spectrum, Ann. Inst. Fourier {\bf 54} (2004)
787--830.

\bibitem{JSS} S. Jitomirskaya, H. Schulz-Baldes  and G. Stolz,
Delocalization in Random Polymer Models, Commun. Math. Phys. {\bf
233} (2003) 27--48.

\bibitem{KKL} R. Killip, A. Kiselev and Y. Last, Dynamical Upper Bounds
on Wavepacket Spreading, Amer. J. Math. {\bf 125} (2003)
1165--1198.

\bibitem{R} C. L. Roy, Some Special Features of Relativistic
Tunnelling through Multi-Barrier Systems with $\delta-$function
Barriers, Phys. Lett. A {\bf 189} (1994) 345--350.

\bibitem{RBa} C. L. Roy and C. Basu, Relativistic Study of
Electrical Conduction in Disordered Systems, Phys. Rev. B {\bf 45}
(1992) 14293--14301.

\bibitem{Si} B. Simon, Schr\"odinger Semigroups, Bull. Amer. Math. Soc.
{\bf 7} (1982) 447--526.

\bibitem{Si2} B. Simon, Absence of Ballistic Motion,  Commun. Math. Phys. {\bf 134}
(1990) 209--212.

\bibitem{SE}C. M. Soukoulis and E. N. Economou, Off-diagonal disorder in one-dimensional systems,
Phys. Rev. {\bf B 24} (1981) 5698--5702.

\bibitem{Sta} A. A. Stahlhofen, Supertransparent Potentials for
the Dirac Equation, J. Phys. A {\bf 27} (1994) 8279--8290.

\bibitem{T} B. Thaller, The Dirac Equation, Springer-Verlag, Berlin,
1991.

\bibitem{TC}G. Theodorou and M. H. Cohen, Extended States In an One-Dimensional System With
Off-Diagonal Disorder, Phys. Rev. {\bf B 13} (1976) 4597--4601.

\bibitem{DrK} H. von Dreifus and A. Klein, A New Proof of Localization
in the Anderson Tight Binding Model, Commun. Math. Phys. {\bf 124}
(1989) 285--299.



\end{thebibliography}
\end{document}